\documentclass{egpubl}
 
\JournalPaper         
\usepackage[T1]{fontenc}
\usepackage{dfadobe}  
\usepackage{amsmath}
\usepackage{amssymb}
\usepackage{cite}  
\BibtexOrBiblatex
\electronicVersion
\PrintedOrElectronic
\ifpdf \usepackage[pdftex]{graphicx} \pdfcompresslevel=9
\else \usepackage[dvips]{graphicx} \fi

\usepackage{egweblnk}
\usepackage{subcaption}

\DeclareMathOperator{\render}{ren}
\DeclareMathOperator{\invrender}{invr}

\makeatletter
\def\thickhline{%
	\noalign{\ifnum0=`}\fi\hrule \@height \thickarrayrulewidth \futurelet
	\reserved@a\@xthickhline}
\def\@xthickhline{\ifx\reserved@a\thickhline
	\vskip\doublerulesep
	\vskip-\thickarrayrulewidth
	\fi
	\ifnum0=`{\fi}}
\makeatother

\newlength{\thickarrayrulewidth}
\title%
     {Half-body Portrait Relighting with Overcomplete Lighting Representation}

\author[Guoxian Song, Tat-Jen Cham, Jianfei Cai, \& Jianmin Zheng]
{\parbox{\textwidth}{\centering Guoxian Song$^{1}$\orcid{0000-0002-3664-572X},Tat-Jen Cham$^{1}$\orcid{ 0000-0001-5264-2572},  Jianfei Cai$^{1,2}$\orcid{0000-0002-9444-3763}
        and Jianmin Zheng$^{1}$\orcid{0000-0002-5062-6226} 
        }
        \\
{\parbox{\textwidth}{\centering $^1$Nanyang Technological University, Singapore\\
		guoxian001@e.ntu.edu.sg, ASTJCham@ntu.edu.sg, ASJMZheng@ntu.edu.sg \\
         $^2$Monash University, Australia\\
         jianfei.cai@monash.edu\\
       }
}
}

\begin{document}
	
	\teaser{
	\includegraphics[width=\textwidth]{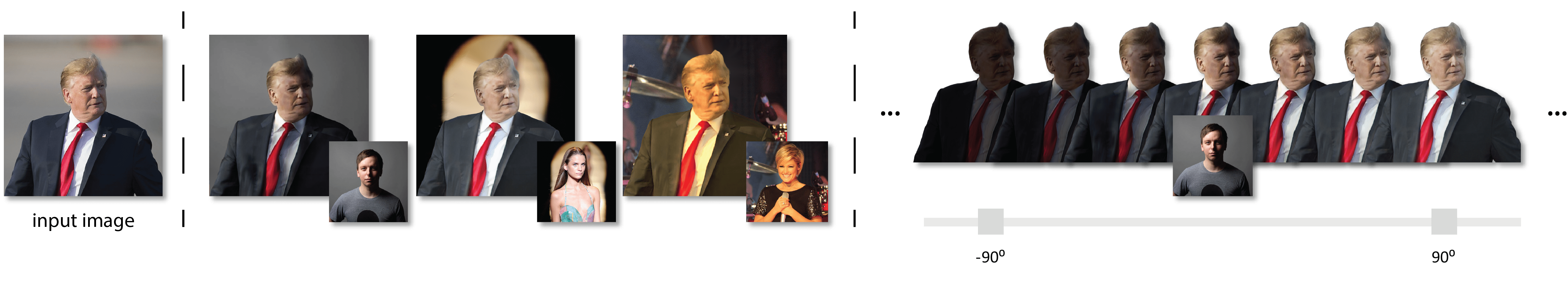}
	  \centering
	   \caption{Given an image, we can generate several relighted images referring to other portrait images. Besides, we can also horizontally rotate the illumination with arbitrary angle without further input.}
		\label{fig:teaser}
	}
	
	\maketitle
	\begin{abstract}
		We present a neural-based model for relighting a half-body portrait image by simply referring to another portrait image with the desired lighting condition. Rather than following classical inverse rendering methodology that involves estimating normals, albedo and environment maps, we implicitly encode the subject and lighting in a latent space, and use these latent codes to generate relighted images by neural rendering. A key technical innovation is the use of a novel overcomplete lighting representation, which facilitates lighting interpolation in the latent space, as well as helping regularize the self-organization of the lighting latent space during training. In addition, we propose a novel multiplicative neural render that more effectively combines the subject and lighting latent codes for rendering. We also created a large-scale photorealistic rendered relighting dataset for training, which allows our model to generalize well to real images. Extensive experiments demonstrate that our system not only outperforms existing methods for referral-based portrait relighting, but also has the capability generate sequences of relighted images via lighting rotations.


	\end{abstract}  
	\section{Introduction}

Portrait relighting is a much sought-after advanced capability in digital photography. While professional photographers may take the time to set up perfect portrait shots meticulously with special equipment, normal users often only care about improving protraits retrospectively, e.g.\ selfie beautification, for which relighting can often help. Commercial applications such as film editing, telepresence and augmented reality can also benefit from this.

In the film industry, highly realistic relighting for special effects may be achieved by first capturing and modeling high quality geometry and reflectances of the subject under heavily instrumented and controlled studio conditions, e.g.\ using a 3D scanner and camera arrays such as the Light Stage\cite{Ghosh:2011:MFC:2070781.2024163,graham2013measurement}. However, such a setup is complex and expensive, thus irrelevant for normal consumers.

Our intention is to have a system that can relight a casually taken portrait image by simply \emph{referring} to another image with the desired lighting condition, and can even interactively adjust the lighting. There have been some advances on the topics related to this goal. In particular, \cite{Shih2014,Shu2017} introduced histogram transfer methods that swap local color statistics between two facial images. However, the visual results are poor when the two images do not share similar facial attributes. Inverse rendering techniques \cite{5279762,DBLP:journals/corr/TewariZK0BPT17,DBLP:journals/corr/abs-1708-00980,Sengupta2018} are able, to some extent, to recover geometry, reflectance and lighting from a single image, but are usually based on the assumption of Lambertian reflectance and a limited second order spherical harmonic (SH) illumination model for the environment. Another option is to directly estimate an environment map from a portrait image~\cite{Sun2019}, which requires a light stage for training data.

\textcolor{black}{Image-based relighting remains an open problem, since it is extremely hard to accurately estimate explicit face geometry, reflectance map and environment map from a single image, for which ground truth is also difficult to obtain. Moreover, incorrect estimates will propagate into the relighting output,  leading to artifacts and unrealistic rendering. There is often insufficient information for these physical attributes to be separable, which makes attempts to estimate them ill-posed. In contrast, we postulate that not all such attributes need to be recovered for relighting purposes; these are only required if conventional (physics-based) rendering has to be used subsequently. To this end, we propose a deep relighting model for half-body portraits, via an implicit form of inverse rendering for portrait relighting and illumination manipulation (see fig.~\ref{fig:teaser}). Unlike previous methods, which explicitly estimate normals, albedo, SH lighting \cite{Sengupta2018} and environment maps \cite{Sun2019}, we implicitly encode the subject and lighting in a latent space, and use the latent codes to generate relighted images by neural rendering. One innovation that we introduce is an overcomplete lighting representation, which facilitates lighting interpolation in the latent space, as well as regularizes the self-organization of the lighting latent space. Moreover, we propose a novel multiplicative neural render for combining subjects and lighting latent codes, leading to improved results.} One unique feature of our proposed neural relighting framework is that we can generate a sequence of relighted images via lighting rotations, based on only a single pair of source and target portrait images.

For the end-to-end training of the proposed neural relighting model, we need a large number of portrait images with multiple known illumination conditions. To the best of our knowledge, there is no such publicly available dataset. Using collected high-quality 3D scans of real humans and real high dynamic range (HDR) environment maps, we rendered numerous photorealistic images to create such a synthetic dataset.

In summary, the main contributions of this paper are:
\begin{itemize}
\item We constructed a large-scale photorealistically rendered dataset using real 3D human scans and HDR environment maps, with lighting rotation annotations for relighting task. 

\item We present a neural relighting framework for half-body portraits, which performs implicit forward and inverse neural rendering. The framework has multiple novel components: the overcomplete lighting representation, the multiplicative neural render and the separated foreground-background lighting encoding. We will release our dataset and code on publication. \footnote{https://github.com/GuoxianSong/Portrait-Relighting.git}

\item We conducted extensive experiments on synthetic and real images. Both qualitative and quantitative results demonstrate that our method outperforms existing methods for referral-based portrait relighting. Moreover, our method can generate a sequence of relighted images for arbitrary horizontal lighting rotations, without need for further input.
\end{itemize}

\section{Related Work}
\subsection{Portrait image capture}
\textcolor{black}{Portrait image can be captured in a professional studio with special equipment such as flashes, deflectors and diffusion panels. In the computer graphics community, other specialized hardware used include 3D scanners and the Light Stage \cite{Debevec:2000:ARF:344779.344855,Ghosh:2011:MFC:2070781.2024163,graham2013measurement} to obtain geometry and reflectance maps, so that photorealistic portrait images can be ray-trace rendered. To capture a large real relighting dataset is challenging, as subjects must remain stationary while the lighting is varied. In our work, we collected high-quality 3D geometry and reflectance maps of real people from artist assets and used a professional rendering engine to synthesize various half-body portrait images.}

\subsection{Image-based relighting}
\textcolor{black}{The histogram based method~\cite{Shih2014} transfers local contrasts and overall lighting from one portrait to another. However, it requires both input and reference pictures with compatible appearance attributes like beards and skin color; otherwise it would generate visible artifacts. Later on, a geometry-aware relighting method~\cite{Shu2017} based on 3DMM (3D morphable models) is able to generate more robust color remapping, but only works with facial regions instead of half-body portraits.}

Inverse rendering  methods\textcolor{black}{\cite{8237510,10.1111/cgf.13220}} aim to decompose images into physical attributes of geometry, reflectance and lighting. This approach has received extended attention over time, especially for human facial images~\cite{Duchene:2015:MII:2843519.2756549,yu2019inverserendernet,Barron:2016:ISP:2914184.2914352}. However, inverse rendering is a difficult ill-posed problem without additional constraint assumptions.
Leveraging on large internet-scale datasets, recent deep learning-based methods such as Mofa~\cite{DBLP:journals/corr/TewariZK0BPT17}, SfSNet~\cite{Sengupta2018}, and \cite{DBLP:journals/corr/abs-1708-00980} have been proposed, which typically use second-order spherical harmonic (SH) functions to model environment lighting, and 3DMM-based facial models to support the estimation of facial geometry, normals and albedo. To make inverse rendering more tractable, these methods typically employ low-frequency illumination models and low-dimensional 3DMM face models. Face-based methods are also limited to relighting facial regions and cannot trivially be extended to half body portraits with visible upper bodies and clothing.

For more general photographs, Zhou et al.~\cite{Zhou} warped 3DMM normals to head regions with an as-rigid-as-possible mapping. They constructed a large facial relighting synthetic dataset using varied SH lighting and proposed a deep single image portrait relighting method. However, their method and dataset only relights the L-channel in Lab color space instead of full RGB. Sun et al.~\cite{Sun2019} collected a `one-light-at-a-time' (OLAT) real human portrait dataset and trained a U-Net based PR-Net to directly estimate environment map, which also facilitates relighting from a single human portrait. However, training this network requires extensive illumination maps and corresponding solid angle supervision, obtained from the physical collection of real images under finely-calibrated ground truth lighting. Broadly, all these methods require extensive optimization or a complex capture setup in order to obtain explicit labels like SH lighting or environment maps. In contrast, our work learns an implicit self-organized overcomplete lighting representation from images, and does not require other additional direct supervisions such as environment maps, SH lighting, geometric normals, etc.

\subsection{Deep image translation}
The seminal work for deep image translation is pix2pix~\cite{isola2017image}, which translates an image from one domain to another using paired training data. It has been applied to many tasks such as inpainting~\cite{pathak2016context}, semantic labeling~\cite{DBLP:journals/corr/DongYWG17} and super resolution~\cite{ledig2017photo}. It has also been extended to unpaired settings~\cite{zheng2018t2net} and for multimodal output~\cite{Zhu2017,Huang2018}. Unsurprisingly, deep image translation has also been used in the task of neural rendering. Thies et al.~\cite{thies2019deferred} proposed a deferred neural rendering method to synthesize images from imperfect 3D content. Meshry et al.~\cite{Meshry} extracted appearance vectors from a deep buffer and injected the vectors into a latent space for neural rendering in-the-wild architectural images. Sengupta et al.~\cite{Sengupta} used inverse rendering and a residual appearance render network for scene images. Inspired by these works, we cast relighting conceptually as a multimodal image synthesis problem, while our network design is also influenced by ideas from classical rendering and inverse rendering processes.

\section{Framework}
\begin{figure*}[t]
	\centering
	\includegraphics[width=1\linewidth]{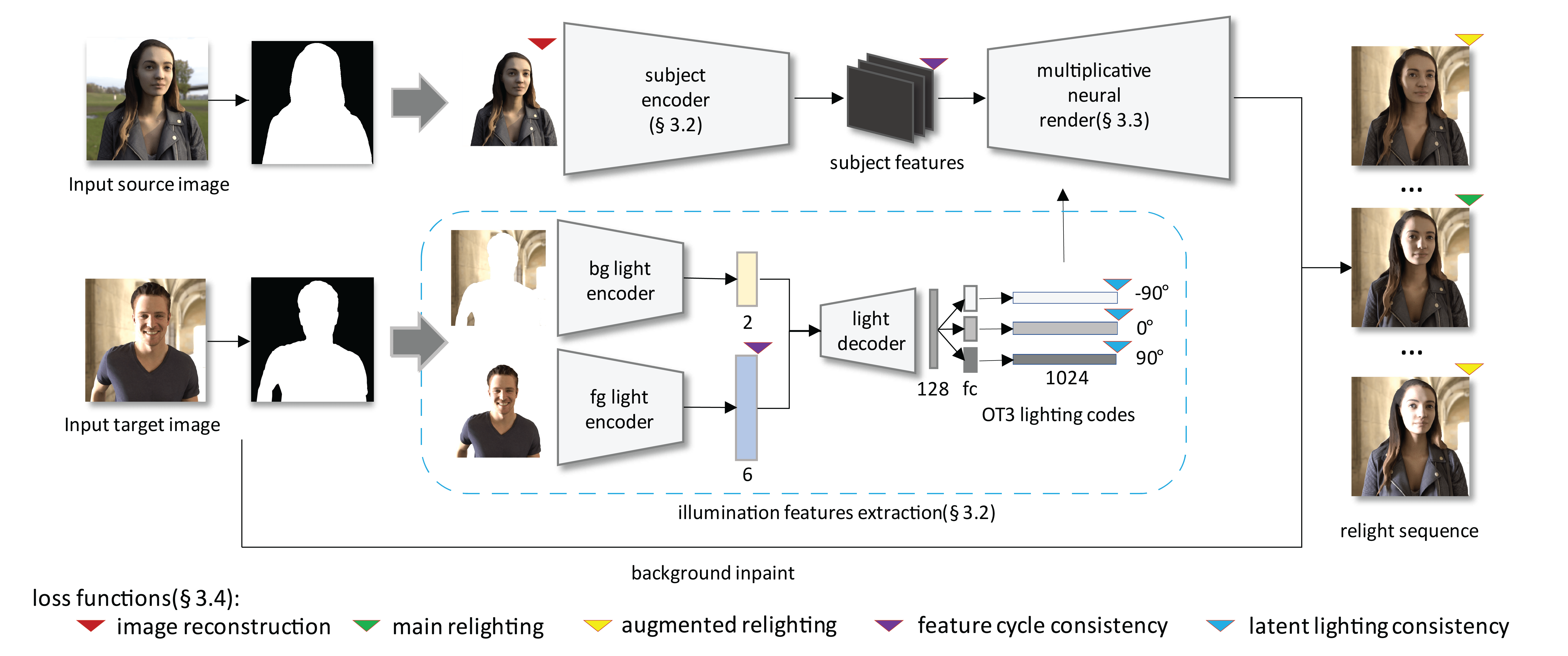}
	\caption{An overview of our portrait relighting system. The locations of loss functions are shown; see text for details.}
	\label{fig:overview}
\end{figure*}

We begin with a simplified formulation for rendering and inverse rendering:
\begin{equation}
	I = \render (s, l); \quad s = \invrender_s(I),\; l = \invrender_l(I)
\end{equation}
where $I$ is an image, $s$ is the subject content, and $l$ is the environment lighting, while $\render(\cdot)$ and $\invrender(\cdot)$ are the forward rendering and inverse rendering functions, respectively. In this framework, $s$ and $l$ remain in latent representations and are not explicitly recovered, and the partitioning of the inverse rendering into separate functions for recovering the subject and lighting features reflects our intention to disentangle these representations.

The focus of this work however is on \emph{relighting} rather than inverse rendering. Given separate scenes $x$ and $y$, the goal is to obtain
\begin{equation}
	I_{x,y} = \render(\invrender_s(I_x), \invrender_l(I_y))
\end{equation}
where $I_{x,y}$ refers to an image in which the subject comes from scene $x$, but is lit by the illumination from scene $y$.

Instead of using only a single latent vector for representing the environment lighting, we propose a new idea of using a \textbf{three-way overcomplete tensorial representation} (denoted \textbf{OT3}):
\begin{equation}\label{eq:ot3}
	\mathbf{L} =
	\begin{bmatrix}
	l^{90} & l^0 & l^{-90}
	\end{bmatrix}
	= \invrender^{\mathrm{OT3}}_l(I)
\end{equation}
where $l^d$ denotes a latent lighting vector corresponding to an environment map that has undergone $d^\circ$ rotation about the vertical axis. In essence, this representation includes extra estimates of the lighting codes should the environmental map be rotated through +$90^\circ$ and -$90^\circ$.  Notice that this representation is \emph{overcomplete}, because $l^{90}$ and $l^{-90}$ do not in fact contain new information not already present in $l^{0}$.

There are two key benefits to this representation:\\
\textbf{1)} The lighting codes are in latent space, and cannot be conventionally manipulated (e.g. via rotation matrices). The three vectors serve as \emph{anchor points} in latent space, facilitating interpolation between them.\\
\textbf{2)} During training, the tensorial representation $\mathbf{L}$ enhances self-organization, because \emph{rotations in environmental lighting will ideally induce column-wise coordinate shifts in the tensor} (akin to pixel-wise translation in images). For example, if the environmental lighting was rotated by $90^\circ$ to form image $I^{90}$, then we should get
\begin{equation*}
	\mathbf{L}^{90} =
	\begin{bmatrix}
	l^{180} & l^{90} & l^{0}
	\end{bmatrix}
	= \invrender^{\mathrm{OT3}}_l(I^{90})
\end{equation*}
and w.r.t.\ (\ref{eq:ot3}) we see that $l^{90}$ and $l^{0}$ have been right-shifted in the tensor. Thus the redundant over-representation can serve the purpose of regularizing the self-organization of the latent space. 


Fig.~\ref{fig:overview} provides an overview of our system. Given source and target images, we extract foreground (human) segmentation maps. A subject encoder $E_s$ is used to extract subject information from the foreground of the source image. Two encoders $E_b$ and $E_f$ are respectively used to extract background and foreground illumination features from the target image. These foreground and background illumination features are concatenated and transferred into the lighting decoder $D$, which generates the three anchor lighting codes in (\ref{eq:ot3}). After that, the subject features are combined with the lighting codes to generate multiple relighted images via the neural render network $R$. Finally, we carry out post-processing by inpainting parts of the target background (de-occluded by removal of subject) and combining with the relighted source foreground.

\subsection{Dataset Generation}

We collected 98 3D scans of real human subjects with highly detailed meshes and high resolution appearance maps from 3D collection websites~\cite{Gobotree,3DScanStore}. We randomly split the dataset into training and test sets of 68 and 30 meshes, respectively. We also collected 228 real outdoor environment maps with 1k resolution from HDRI Haven~\cite{HDRIHaven}, in which training and test sets were randomly split into 160 and 68 maps respectively. We used a professional ray-tracing engine \cite{ArnoldRenderer} to render photorealistic images.

\begin{figure}[h]
	\includegraphics[width=1\linewidth]{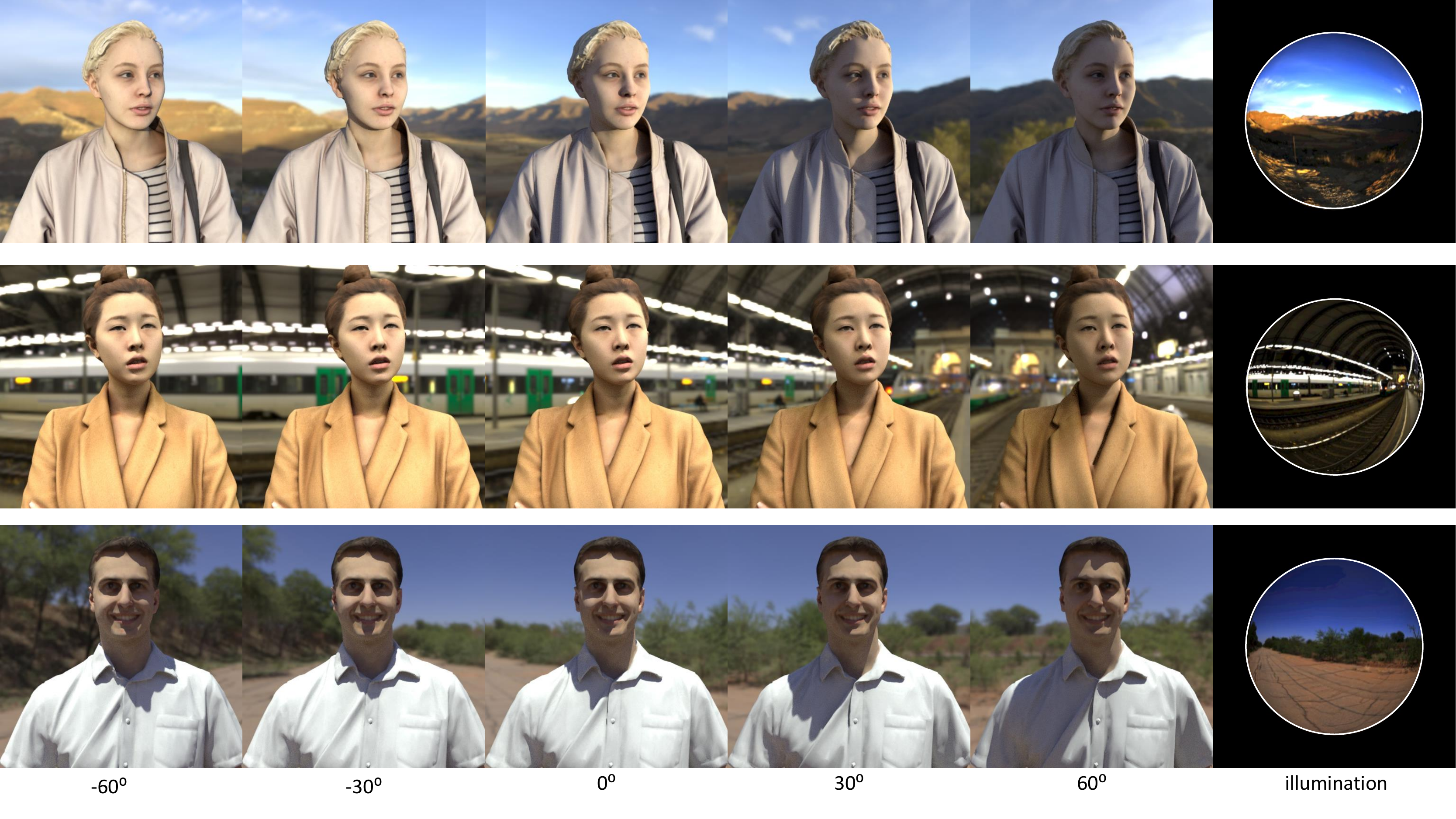}
	\caption{Samples from our dataset.}
	\label{fig:dataset}
\end{figure}

We rendered half-body figures from viewpoints which are front-facing or partly oblique. First, we coarsely aligned the figures via global rotation, scaling and translation, such that the 3D torsos had similar widths. Then for each environment map, we created a series of scenes in which the environment illumination was sequentially rotated through $30^{\circ}$ around the vertical axis 12 times. Finally, we rendered images and foreground segmentation maps, based on a camera with a 45 mm focal length and from a fixed viewpoint. The size of the images are $512\times512$, suitable for network input. Examples can be seen in fig.~\ref{fig:dataset}. In total, we generated 130,560 training images under 1,920 illuminations and 24,480 testing images under 816 illuminations, with no further augmentation. Note that there is no overlap between training and test subjects, nor between training and test illuminations. Furthermore, training illuminations are only used with training subjects, and likewise for the test images.

\subsection{Illumination and Subject Encoding}

In a portrait image, the appearance of the subject is mainly modulated by the foreground lighting, i.e. the illumination behind the camera, but this is hard to estimate (hence techniques such as~\cite{LeGendre2019}). The background lighting behind the person can be directly observed, but only has an indirect effect on the appearance; nonetheless there may be significant correlation to the foreground lighting, e.g. in a mountain scene, the background grass and lake may help estimate the dominant sun direction behind the camera.

In our framework, we initially have separate foreground and background lighting encoders of different channel sizes, with foreground lighting having more channels due to its greater significance. These separate illumination features are subsequently concatenated and further decoded into the three-way overcomplete latent representation of the lighting, for rendering the subject.


Specifically, given an input image $I$ with subject segmentation mask $M$, our illumination encoders $E_b$ and $E_f$ extract features $i = (i_b, i_f) = (E_b( I_b), E_f(I_f))$ from the foreground $I_f = I \odot M$ and background $ I_b = I \odot (1- M)$ image regions. As we do not intend in this work to inversely render a detailed environment map, the encoders are highly compressive, with $i_f$ a 6-dim vector and $i_b$ a 2-dim vector. Next, $i$ is expanded via a common multi-layer perceptron and three separate fully-connected (fc) layers to obtain the three lighting codes $\{l^{90}, l^0, l^{-90}\}$, each a 1024-vector.

In parallel to the above, we also need to encode the subject in the source image. The source image and the foreground segmentation map are passed through a subject encoder $s = E_s(I_f)$ to obtain the subject features $s$. The features $s$ and $i_f$ are disentangled due to the loss functions used in training (later in sec.\ \ref{sec:loss}).


\subsection{Multiplicative Neural Rendering}
\begin{figure}[t]
	\includegraphics[width=1\linewidth]{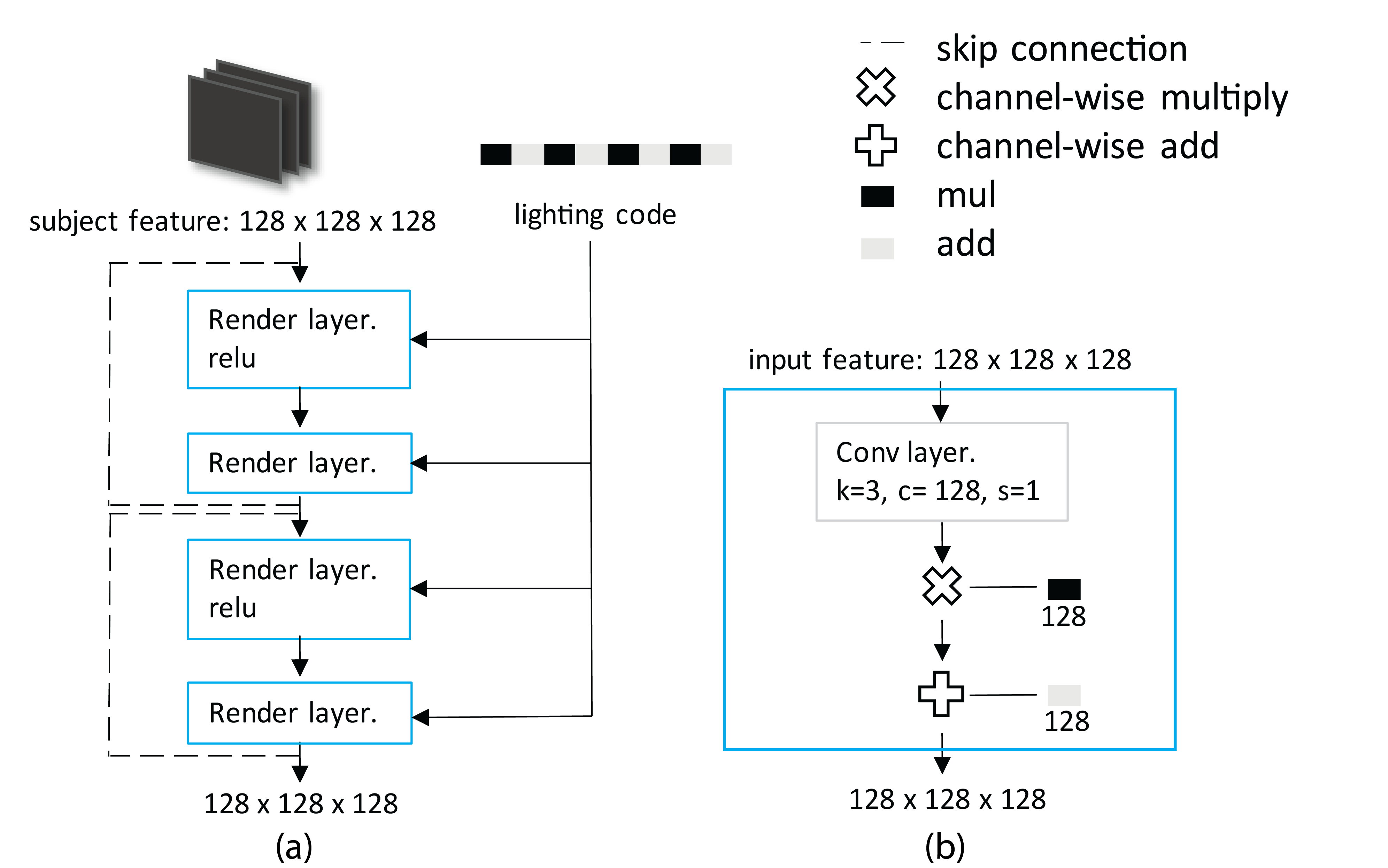}
	\caption{Illustration: (a) multi-scale multiplicative neural render; (b) multiplicative render layer.}
	\label{fig:MNR}
\end{figure}
After obtaining the lighting codes and the subject features, the subsequent step is to render the image, which has to be done neurally since the representations are all latent in nature. A straightforward way is to simply concatenate lighting and subject features and decode into image space~\cite{Meshry}. However, from our investigations this does not easily lead to good results\textcolor{black}{(See Sec.4.2.2.)}. A possible reason is that a deconvolutional network may not have the ideal structural form to model the rendering process from this representation.

Our approach is to adopt some structural aspects from classical graphics rendering models. These typically involve the product of illumination and reflectances, with an additive component of ambient illumination. As such, instead of simply concatenating lighting and subject features, a combination of multiplication and addition is used to combine the lighting codes and subject features, which we call \emph{multiplicative neural render} (MNR). While there is no direct correspondence between multiplication in classic rendering and neural multiplication of latent features, empirically we found that this works well compared to concatenation or purely addition.

\textcolor{black}{First, each lighting vector $l \in \mathbb{R}^{1024}$ is sequentially partitioned into 4 pairs of lighting sub-codes representing multi-scale multiplicative and additive components, i.e. $l=(l^1_{mul}, l^1_{add}, l^2_{mul}, l^2_{add}, \ldots)$, where the superscript denotes the corresponding render layer index, and each component is in 128 dimensions, as illustrated in Fig.~\ref{fig:MNR}(a).} For the first render layer, the following neural multiplication takes place:
\begin{equation}
	out^1(l, s) = l^1_{mul}\times s + l^1_{add} .
\end{equation}
The output of the render layer $out^1(l, s)$ is then used as the input for the next render layer, illustrated in Fig.~\ref{fig:MNR}(a). The operation is performed in the channel dimension (Fig.~\ref{fig:MNR}(b)), which keeps our network fully convolutional. \textcolor{black}{We stack four multiplicative neural render layers to relight deep subject features, and the lighting vector progressively changes the subject features. }

After relighting the foreground subject, we can further embed the subject into the background of the target image. To cope with the removal of the foreground in the target image, we first use a simple fast marching approach \cite{Telea04} to inpaint and complete the target image background, after which the source foreground subject can be composited.

\begin{figure*}[hbt!]
	\centering
	\includegraphics[width=0.93\linewidth]{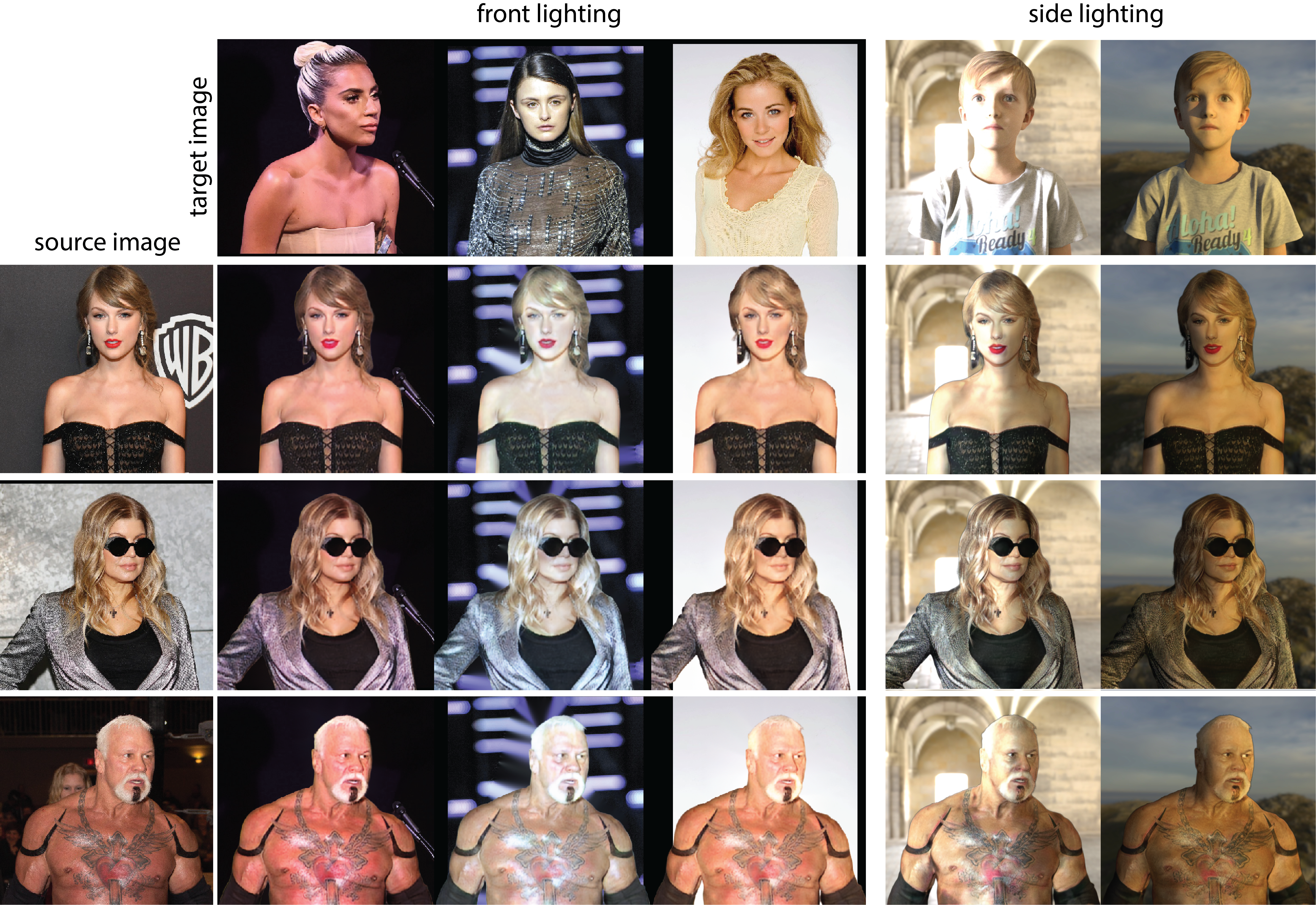}
	\caption{\textcolor{black}{Our relighting results of half body portraits of real images from CelebA dataset~\cite{liu2015faceattributes}. The first col are source images, and the first row are target images for two lighting conditions: front lighting(col. 2-4) and side lighting(col. 5-6). The front lighting images are from CelebA dataset, and side lighting images are from our test dataset.  }}
	\label{fig:our_real_result}
\end{figure*}

\subsection{Loss functions}
\label{sec:loss}

To train the networks, we use several loss functions. The parts of our system to which these loss functions apply are indicated in Fig.~\ref{fig:overview}. \\[3pt]
\textbf{Image reconstruction:} The neural renderer $R$ should reconstruct the foreground subject in the source image ${I_x}$, from extracted subject feature $s_x$ and lighting code $l_x^{0}$:
\begin{equation}
	L^x_{recon} = || M_x \odot (R(l_x^{0}, s_x) - I_x)||_{1}
\end{equation}
\textbf{Main relighting:} Given $I_x$ and lighting code $l_y^{0}$ from $I_y$, the relighted image should be close to the groundtruth $I^{0}_{x,y}$:
\begin{equation}
	L^{x,y}_{relight} = || M_x \odot (R(l_y^{0}, s_x) - I^0_{x,y})||_{1}
\end{equation}
\textbf{Augmented relighting:} Since the OT3 lighting representation provides two other lighting codes $l^{-90}_y$ and $l^{90}_y$, we can render additional relighted images for comparing to the groundtruth $I^{-90}_{x,y}$ and $I^{90}_{x,y}$:
\begin{multline}
L_{auglight}^{x,y} = ||  M_x \odot (R( l^{90}_{y} ,s_x) - I^{90}_{x,y} )||_{1} + \\
||  M_x \odot (R( l^{-90}_{y} ,s_x) - I^{-90}_{x,y} )||_{1}
\end{multline}
\textbf{Feature cycle consistency:} Given illumination features $i = (i_b, i_f)$ randomly sampled from normal distribution and subject features $s_x$ from $I_x$, we can render a relighted image $\hat{I}_{x,y} = M_x \odot R(l^0, s_x)$, where $l^0 = D^{0}(i)$ is the $0^{\circ}$ lighting code mapped from $i$ by the MLP decoder $D^0$. If $\hat{I}_{x,y}$ is re-encoded again, we want the recomputed subject and foreground-only illumination features to be consistent with $s_x$ and $i_f$ respectively. The random noise greatly helps data augmentation and encourages $s_x$ and $i_f$ to be disentangled, which benefits from implicit representation.
\begin{equation}
	L_{feat}^{x,y} = ||E_s(\hat{I}_{x,y}) - s_x ||_{1} + ||E_f(\hat{I}_{x,y}) - i_f||_{1}
\end{equation}
\textbf{Latent lighting consistency:} Given image $I_y$ and lighting-rotated counterparts $I^{-90}_y$ and $I^{90}_y$, we can extract the overcomplete OT3 lighting codes from each. Suppose the codes for $I_y$ are $\{l_y^{-90}, l_y^{0}, l_y^{90}\}$. Considering that the codes for $I^{-90}_y$ and $I^{90}_y$ overlap with $I_y$'s at four instances, and overlap with each other's at the lighting rotation of $\pm 180^\circ$, we add the following latent lighting consistency loss by denoting $E_i$ as the concatenated form of $E_b$ and $E_f$:
\begin{multline}
\hspace*{-10pt}
L_{cons}^{y} =
||l_y^{90} - D^0( E_i(I^{90}_y)   ) ||_{1} + || l_y^{0} - D^{-90}(E_i(I^{90}_y)) ||_{1}+\\
||l_y^{0} - D^{90}(E_i(I^{-90}_y)) ||_{1} + || l_y^{-90} - D^{0}(E_i(I^{-90}_y)) ||_{1} + \\
|| D^{-90}(E_i(I^{-90}_y))- D^{90}(E_i(I^{90}_y)) ||_{1}
\end{multline}
\textbf{Total loss:} We jointly train the encoders ($E_s$, $E_f$ and $E_b$), MLP lighting decoder $D$ and neural renderer $R$ to optimize the combined objective function:
\begin{multline}
	L_{total}^{x,y} = L^x_{recon} + L^{x,y}_{relight} +
	 \lambda_a L_{auglight}^{x,y} + \lambda_f L_{feat}^{x,y} + \lambda_c L_{cons}^{y}
\end{multline}
where $\lambda_a$, $\lambda_f$, $\lambda_c$ are set as 0.5, 0.1, 0.25, respectively.

\textcolor{black}{The design of our relighting network facilitates disentangling portrait images into implicit illumination and subject content features. By employing our proposed OT3, the overcompleteness of this representation helps the learned latent space to self-organize in a more coherent manner, that allows for better lighting interpolation. Also of importance to improving neural rendering quality is the use of a novel multiplicative neural render, as is the use of the key loss functions described above that further help to regularize the output. The impact of these contributions will be shown in the next section on experiments.}


\begin{figure*}[htp]
	\centering
	\includegraphics[width=0.95\linewidth]{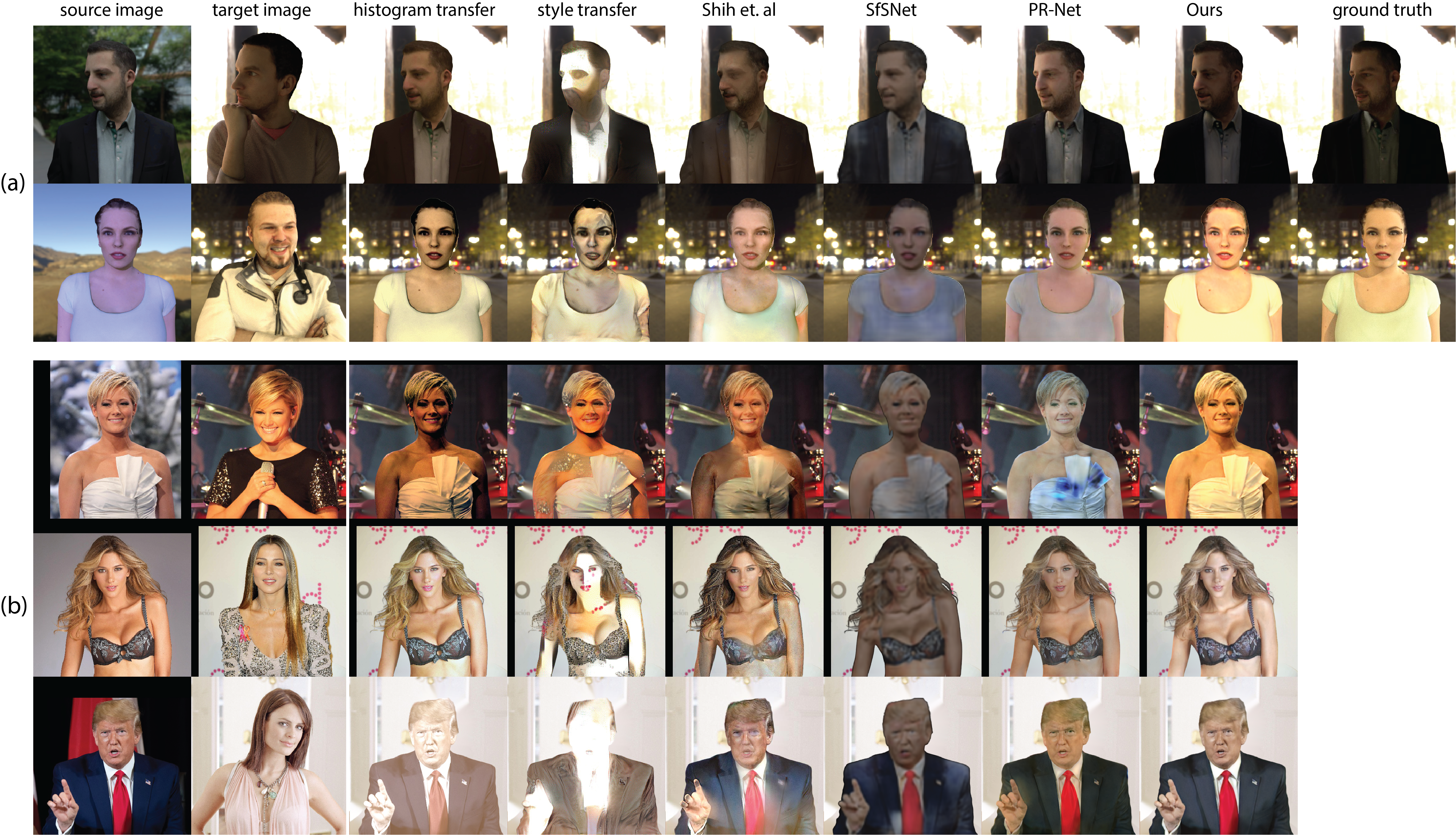}
	\caption{\textcolor{black}{Qualitative comparisons of our method with two classical methods: histogram transfer~\cite{10.1145/1128923.1128974} and style transfer\cite{gatys2015neural}, and three state-of-the-art relighting techniques: Shih et. al\cite{Shih2014}, SfSNet~\cite{Sengupta2018}, PR-Net~\cite{Sun2019} on our synthetic test set in (a) and real images from Celeb A in (b). Note that for real images, there is no ground truth relighted image.}}
	\label{fig:visual_comparson}
\end{figure*}

\section{Experiments}
\begin{figure*}[ht]
	\centering
	\includegraphics[width=0.95\linewidth]{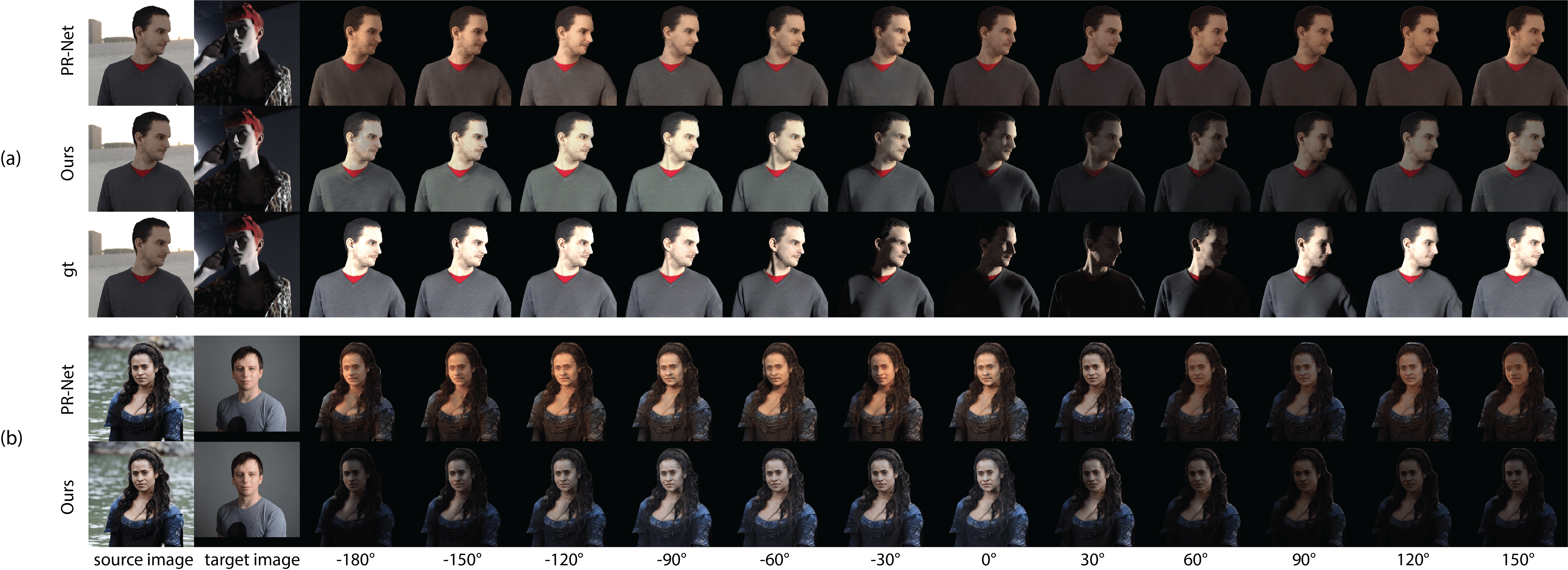}
	\caption{Visual comparison of our method and PR-Net~\cite{Sun2019} of lighting rotation task from $-180^{\circ}$ to $150^{\circ}$ on test set in (a) and real images from Celeb A in (b). For clarity, we exclude the background in the relighting.}
	\label{fig:compare_rotation}
\end{figure*}

\textbf{Implementation}
Our encoders, decoder and neural render networks follow recent autoencoder network structures in \cite{kingma2013autoencoding,Huang2018,Zhu2017,CycleGAN2017} with residual blocks \cite{he2016deep}. The subject encoder down-samples a 512$\times$512$\times$3 image to a $128\times128\times128$ feature with convolutional layers and residual blocks. The two lighting encoders convert a target image into a compressive latent feature with 8 dimensions, using convolutional layers, global average pooling, a fully connected layer and concatenation. The lighting decoder transforms this compressive light feature to three lighting codes of 1024 dimensions each. The neural render network uses the lighting codes to reweigh the subject feature through four neural multiplicative layers in progression, and finally up-samples to a 512$\times$512$\times$3 output using several deconvolutional layers. The network details are in the appendices material.

We jointly trained all the networks using the Adam optimizer with a learning rate of $1.5$$\times$$10^{-5}$, a batch size of 2, and 5 epochs on an NVIDIA Quadro P5000 in Pytorch \cite{paszke2017automatic}.

\begin{figure}[h]
	\centering
	\includegraphics[width=1.0\linewidth]{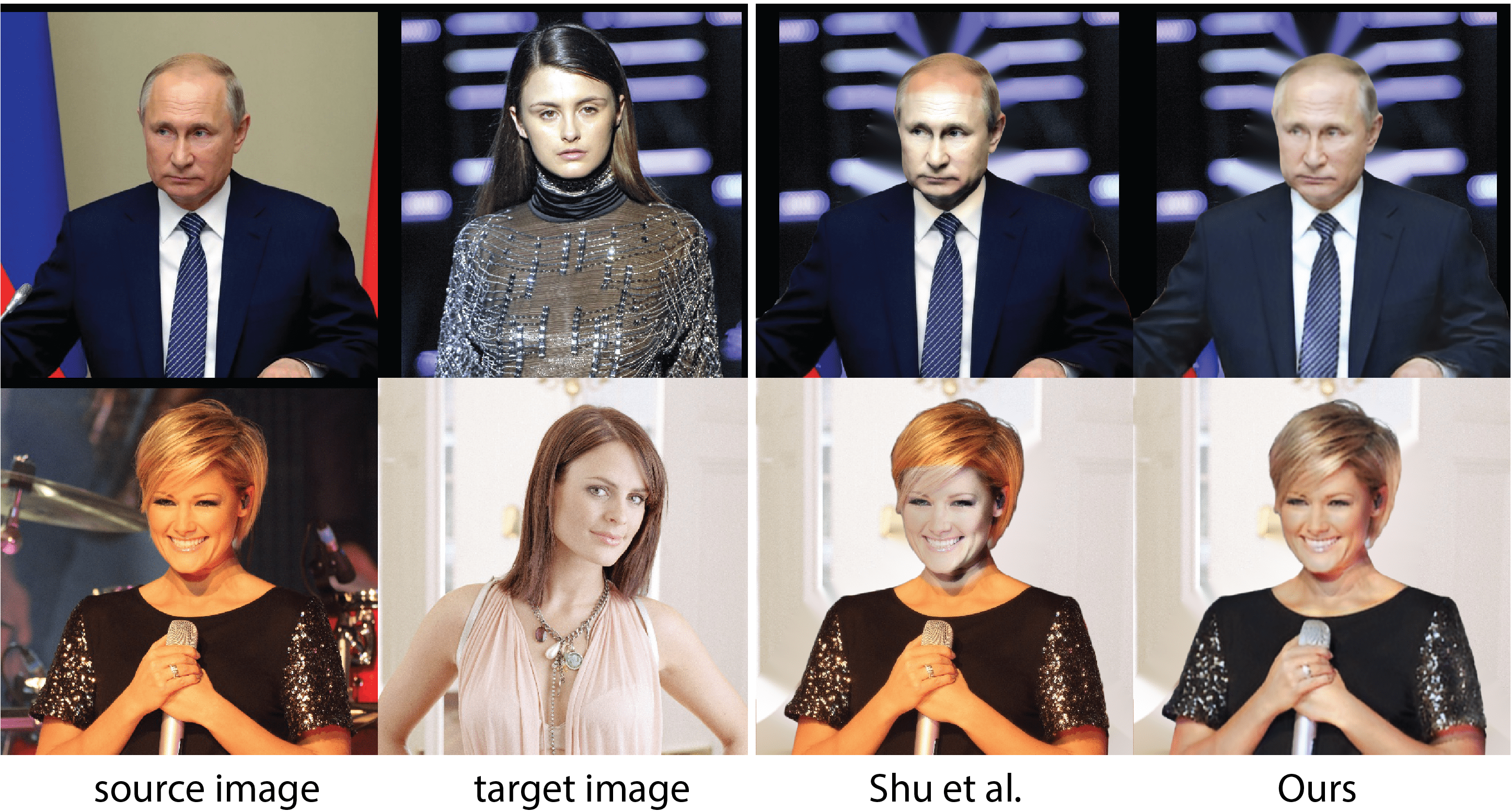}
	\caption{Visual comparisons of our method and the 3D face approach of Shu et al. \cite{Shu2017}.}
	\label{fig:Shu.et_al}
\end{figure}

\subsection{Comparison to Existing Methods}

\subsubsection{Overall visual results}Fig.~\ref{fig:our_real_result} shows our results of relighting real images from the CelebA dataset \cite{liu2015faceattributes}. Real images were pre-processed to segment foreground from background using tool~\cite{remove.bg}. For each input source and target pair, we show the relighted composited image. It can be seen that our method can robustly handle a variety of real photos containing different subjects, illumination conditions and clothing styles, although we trained the networks using only our synthetically rendered images. This suggests that our rendered dataset is sufficiently realistic.

\vspace{-0.15cm}
\subsubsection{Qualitative comparison} \textcolor{black}{In Fig.~\ref{fig:visual_comparson}, we visually compare our method on the single image relighting task with two classical methods: histogram transfer~\cite{10.1145/1128923.1128974} and style transfer~\cite{gatys2015neural}, as well as three state-of-the-art relighting methods: a multi-scale histogram transfer method~\cite{Shih2014}, the intrinsic image decomposition method of SfSNet~\cite{Sengupta2018} and PR-Net~\cite{Sun2019}. Classical histogram transfer~\cite{10.1145/1128923.1128974} matches pixel color distributions while style transfer~\cite{gatys2015neural} matches deep VGG19 features, and in both cases the matching was performed within the portrait regions.  The multi-scale histogram transfer method of \cite{Shih2014} separately matches local low and high frequency statistics of the target portrait regions. As for the other mentioned learning-based methods, they were all re-trained on our training dataset. For SfSNet, we first simulated having a white probe ball in each scene and approximated its second order Spherical harmonic (SH) lighting vectors, which were used in the training. Relighting was performed by rendering, which combines the normals and albedo estimated by the model from the source image, with the estimated SH weights from the target image. Because this model only supports 128×128 resolution input, we upsampled the output images to 512×512. For PR-Net, since their dataset and code were not publicly available, we implemented their Unet-based PR-Net using the settings in their paper. To reproduce their method, we further downsampled 1K environment maps to 32x16 and simulated corresponding solid angle maps in training. Relighting was performed by combining the inferred environment map from the target image and the encoded features from the source image. For fair comparison, each method was applied to relight only the person in the image, which was then composited with the inpainted background. }

\textcolor{black}{From Fig.~\ref{fig:visual_comparson}, it can be seen that our method successfully relighted subjects with perceptually correct lighting from directional illumination and ambient lighting.} The histogram transfer method (column 3) is unable to reproduce geometry-dependent side lighting effects (row 1), and is easily affected by differently colored clothing (rows 3 \& 5). The deep style transfer method (column 4) does not correctly disentangle illumination and portrait content, leading to obvious artifacts. The results of Shih et al.~\cite{Shih2014} suggest that multi-scale histogram transfer is not appropriate for lighting transfer when two images do not share common appearance features. The performance of SfSNet heavily depended on its intrinsic decomposition results (i.e.\ albedo, normal and SH lighting), which can lead to poor relighting results if the decomposition is not accurate. Moreover, the second-order SH used in SfSNet to model lighting did not model complex scene lighting well, as expected. PR-Net~\cite{Sun2019} did not sufficiently relight images, with source illumination appearing to remain the same. This might be due to their U-Net structure, which kept source illumination during image generation. Also, their model struggled when presented with clothing containing highly saturated colors.

\begin{table}
	\centering
 \begin{subtable}{0.5\textwidth}
	\centering
	    \begin{tabular}{c|ccc}
		\thickhline
		Algorithm & RMSE& PSNR &SSIM \\
		\hline\hline
		histogram transfer~\cite{10.1145/1128923.1128974} & 0.196 & 15.23& 0.85 \\
		style transfer~\cite{gatys2015neural} & 0.265 & 11.92& 0.80 \\
		Shih et. al~\cite{Shih2014}  & 0.181 & 15.76& 0.87 \\
		SfSNet~\cite{Sengupta2018}   & 0.170&16.15&0.86 \\
		PR-Net~\cite{Sun2019} & 0.117&19.50&0.88 \\
		\hline
		Ours      & \textbf{0.099}&\textbf{21.08}&\textbf{0.90}\\
		\thickhline
	    \end{tabular}
	    \caption{Single image relighting result.}
    \end{subtable}
    
  \vspace*{0.5 cm}
 \begin{subtable}{0.5\textwidth}
	\centering
		\begin{tabular}{c|ccc}
		\thickhline
		Algorithm & RMSE& PSNR & SSIM   \\
		\hline\hline
		PR-Net~\cite{Sun2019} &0.123&19.15&0.88      \\
		\hline
		Ours      &  \textbf{0.105}&\textbf{20.57}&\textbf{0.89}\\
		\thickhline
    	\end{tabular}
    	\caption{Sequential relighting result.}
	\end{subtable}
	\caption{\textcolor{black}{Quantitative results for comparison. We measure root mean square error (RMSE), peak signal-to-noise ratio (PSNR) and structural similarity (SSIM) on the human regions of about 24k relighted testing images w.r.t. the ground truth. Note that histogram transfer~\cite{10.1145/1128923.1128974}, style transfer~\cite{gatys2015neural},  Shih et al.~\cite{Shih2014} and SfSNet~\cite{Huang2018} cannot be used to generate sequential output. }}
	\label{tab:quantitative}
\end{table}

\begin{figure}[h]
	\includegraphics[width=1.0\linewidth]{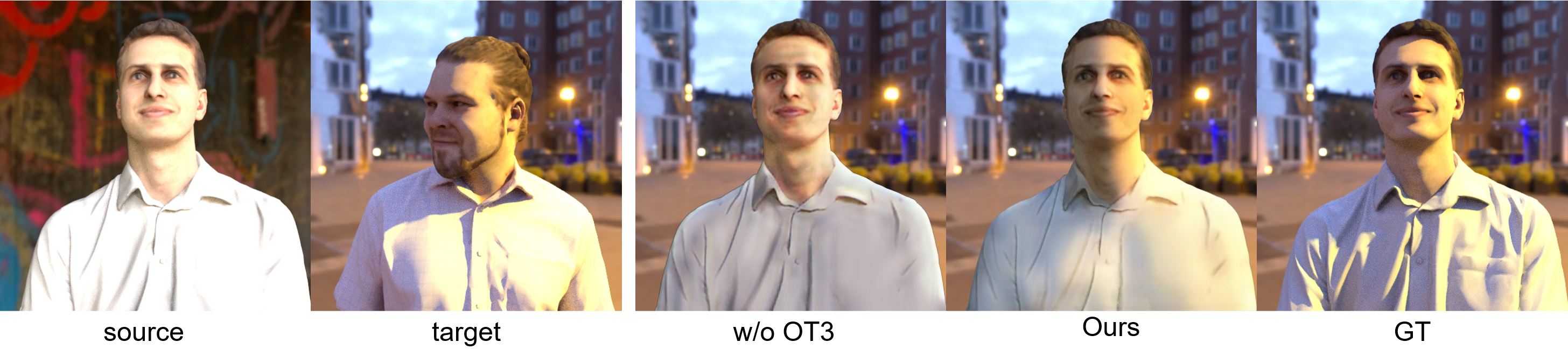}
	\caption{\textcolor{black}{Visual comparison of three-way overcomplete tensorial representation(OT3) with single tensorial representation.}}
	\label{fig:compare_ot3}
\end{figure}

\begin{figure}[h]
	\includegraphics[width=1.0\linewidth]{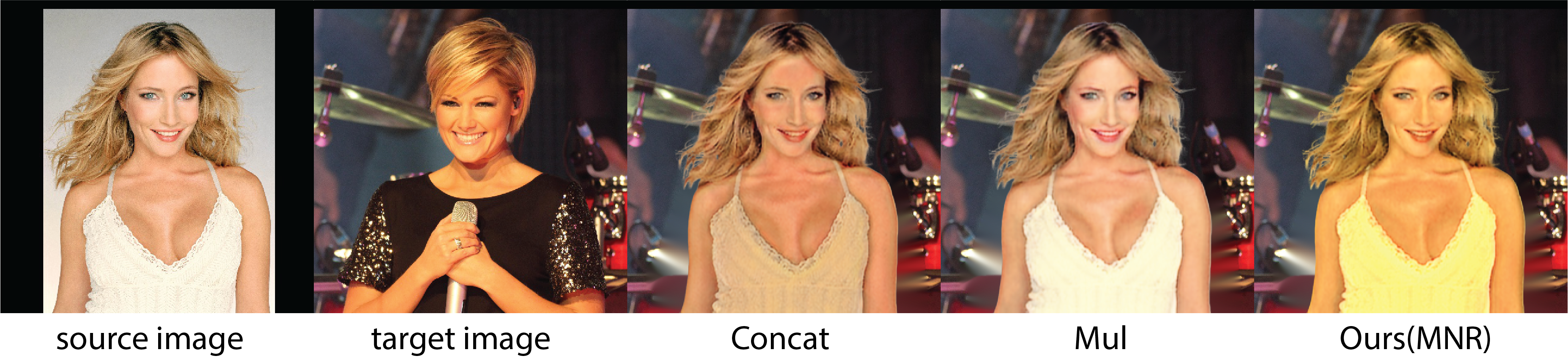}
	\caption{Visual comparison of concatenation-based (Concat), multiplication-based (Concat) vs multiplicative neural render (MNR), both in vector latent space.}
	\label{fig:compare_inject}
\end{figure}

\begin{table}
	\centering
	\begin{tabular}{c|cccc}
		\thickhline
		Measurement &  Concat& Mul & MNR(ours)  \\
		\hline
		RMSE &0.110& 0.108 &\textbf{0.105} \\
		PSNR &20.21& 20.19 &\textbf{20.51}\\
		\thickhline
	\end{tabular}
	
	\caption{\textcolor{black}{Quantitative comparisons of three different neural render approaches on our test dataset.}} 
	\label{tab:quantitative_concat}
\end{table}

\begin{table*}[h]
	\centering
	\begin{tabular}{c|ccccc}
		\thickhline
		Measurement & w/o BG & w/o OT3& w/o $L_{feat}$ & w/o $L_{cons}$ & Full    \\
		\hline
		RMSE &0.104&0.105 &0.178 &0.105  &\textbf{0.099} \\
		PSNR &20.80&20.51&16.24 &20.46 &\textbf{21.08}\\
		\thickhline
	\end{tabular}
	
	\caption{\textcolor{black}{Quantitative results where different components are ablated.}} 
	\label{tab:quantitative_albation}
\end{table*}

\vspace{-0.15cm}
\subsubsection{Quantitative comparison}
Table~\ref{tab:quantitative} compares quantitative results of our method to existing methods. Evaluation was done with the three metrics of RMSE, PSNR and SSIM. For each of the 24,480 images in the test dataset, we randomly chose another image from the entire dataset as the target lighting image. It can be seen that for the relighting results, our method achieved the best performance over all metrics, particularly improving the existing best PSNR result of 19.50 by 1.58 dB. \textcolor{black}{We also compared our method with PR-Net on the sequential outputs by sequentially relighting the same test dataset with multiple $30^{\circ}$ lighting rotation offsets. Specifically, we first extracted lighting codes $\{l^{90}, l^0, l^{-90}\}$ from a target image, and then inversed the fully connected layer for $l^{-90}$ to estimate pseudo $l^{-180}$ anchor lighting code. This enables piecewise linear interpolation to get lighting codes for intermediate angles across the full $360^\circ$ range. We measured the average quantitative results in the three metrics under $\{ -180^{\circ},-150^{\circ},..,120^{\circ}, 150^{\circ}\}$, in total 12 lighting rotations for each test image. From Table~\ref{tab:quantitative} it can be seen that our method outperformed PR-Net, improving PSNR by 1.42 dB for sequential relighting. }

\vspace{-0.15cm}
\subsubsection{Comparison on lighting rotation}
Fig.~\ref{fig:compare_rotation} shows lighting-rotated relighted images of our method and PR-Net~\cite{Sun2019}. \textcolor{black}{Accurately estimating the environment map from a single image is very ill-posed and may be an overkill for this problem. In contrast, our method estimates anchor lighting representations and interpolates lighting codes in the latent space, which lead to more reasonable lighting rotation effects with smooth transitions.}

\vspace{-0.15cm}
\subsubsection{Comparison to the 3D face approach of~\cite{Shu2017}}
We also compared our method with the 3D face approach of Shu et al.~\cite{Shu2017} for single-image relighting. Fig.~\ref{fig:Shu.et_al} shows the visual comparisons on two real-image examples. Shu et al. first estimated 3DMM parameters for the face, and then applied mass transport to the local histogram. Since their approach is mainly based on 3DMM, it is restricted to relighting facial regions and cannot handle the rest of the body. In contrast, our method can relight both the face and upper body. Furthermore, taking observations from a larger region of the image would as expected lead to better inferred lighting. Note that we do not report the quantitative results of \cite{Shu2017} in Table~\ref{tab:quantitative} since it only relights facial regions. In addition, \cite{Shu2017} took 228s to handle each input image pair due to their optimization-based approach, whereas our method only took 0.3s during test time.

\subsection{Ablation Studies}

\subsubsection{Overcomplete Lighting Representation}
\label{sec:ablate_OT3}
\textcolor{black}{We verified the effectiveness of our three-way overcomplete tensorial representation (OT3) by replacing it with a single tensorial representation.
After doing so, we re-trained the network on our relighting dataset. The visual results are shown in Fig.~\ref{fig:compare_ot3}.}

\textcolor{black}{It can be seen that the standard representation with a single tensor is only able to generate mild shading when given the target side lighting, and is far from the ground truth. The is likely because the latent space is not structurally regularized enough to be consistent with the underlying illumination, and has reduced generalizing power. Conversely, the added structural redundancy of OT3 led to a result that is much closer to the ground truth.
The quantitative results in Table~\ref{tab:quantitative_albation} further demonstrate the advantage of our OT3 representation. }

\subsubsection{Multiplicative Neural Rendering}
\textcolor{black}{We verified the effectiveness of our multiplicative neural rendering (MNR) by comparing it with a simple concatenation approach~\cite{Meshry} (denoted "Concat") and \textcolor{black}{directly latent multiplication approach(denoted "Mul")}. For clarity, we excluded the use of OT3 codes in this case, and only compared Concat, Mul and MNR trained on a single-vector latent-space representation. For the Concat, Mul and MNR configurations, the subject and the lighting encoder architectures remain the same, but Concat/Mul directly concatenates or multiply the illumination features to subject features as input to the neural render network. The visual results are shown in Fig.~\ref{fig:compare_inject}, where Concat and Mul produce limited relighting results, while our MNR's output appears perceptually more realistic.} 

\textcolor{black}{In addition, we performed an quantitatively evaluation to compare our MNR with the conventional Concat/Mul method. The result in Table~\ref{tab:quantitative_concat} shows our MNR consistently outperforms Concat and Mul.}

\subsubsection{Quantitative ablation results}
We additionally performed an ablation study of our own model to quantitatively evaluate the individual model components and the introduced additional loss functions: \textcolor{black}{ 1) We used only one lighting encoder to extract 8-dim illumination features from the foreground without the background (w/o BG), to determine the importance of using the portrait background in estimating the illumination, noting that the lighting that affects the portrait predominantly comes from \emph{behind the camera}. 2) We used only one lighting vector code instead of a 3-code tensor (w/o OT3), as per Sec.~\ref{sec:ablate_OT3}. 3) We removed the $L_{feat}$ feature cycle consistency loss during training that would have provided the incentive for the subject encoder and multiplicative neural render networks to be cyclically consistent, when the rendered output is recycled as input. 4) We removed the $L_{cons}$ latent lighting consistency loss during training that would have encouraged the OT3 codes to correctly overlap for $\pm 90^\circ$ lighting-rotated input target images. We retrained the network for each scenario on our training dataset and evaluate on our test dataset. From the results in Table~\ref{tab:quantitative_albation}, we can see that all the proposed components and loss functions contributed to performance improvement. Among these, the feature cycle consistency loss for lighting and content disentanglement appeared to be the most significant for   relighting quality. }

	\begin{flushleft}
		
	\end{flushleft}
	
	\section{Concluding Discussion}
	
	We presented an image-based deep generative model that can dynamically relight half-body portrait images. Key technical contributions include the introduced OT3 lighting representation, the multiplicative neural rendering and the separation of background and foreground for illumination feature encoding. We have also created a large rendered dataset with annotated and controlled lighting that is suitable for training our model, and with sufficient photorealism to allow our model to be directly applied to real images. Extensive qualitative and quantitative results have demonstrated the superior performance of our proposed method on both synthetic and real images.
	
	\begin{figure}
		\centering
		\includegraphics[width=0.80\linewidth]{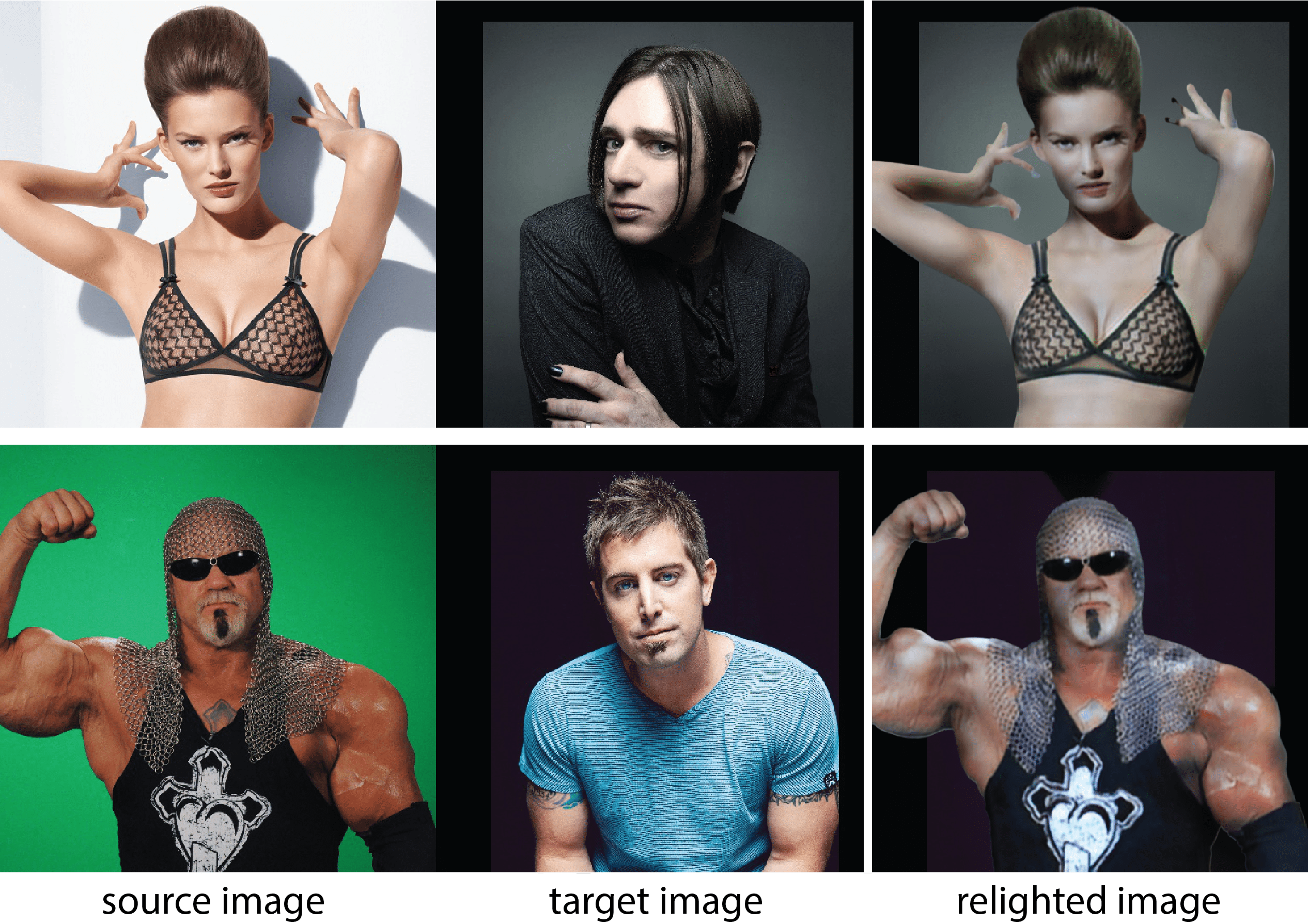}
		\caption{Examples of failure cases due to preserve hard shadowing (neck) and sharp specularities (glasses).}
		\label{fig:limitation}
	\end{figure}
	
	Our method has some limitations. Fig.~\ref{fig:limitation} shows two typical failure cases for our model. These input images contain hard shadowing or sharp specularities.
	A possible solution is to involve direct reasoning in 3D to handle geometric-sensitive shadows and specular reflections. Another possible solution is to augment our training data for these under-represented scenarios.

	\section{Acknowledge}
	This research was conducted at Singtel Cognitive and Artificial Intelligence Lab for Enterprises (SCALE@NTU), which is a collaboration between Singapore Telecommunications Limited (Singtel) and Nanyang Technological University (NTU) that is supported by A*STAR under its Industry Alignment Fund (LOA Award number: I1701E0013).
	

	\bibliographystyle{eg-alpha-doi} 
	\bibliography{egbibsample}       

\newcommand{\etalchar}[1]{$^{#1}$}
\begin{thebibliography}{\uppercase{DYWG17}}

\bibitem[3D ]{3DScanStore}
{3D Scan Store}.
\newblock \url{https://www.3dscanstore.com}.

\bibitem[Arn]{ArnoldRenderer}
{Arnold Renderer}.
\newblock \url{https://www.arnoldrenderer.com}.

\bibitem[BM16]{Barron:2016:ISP:2914184.2914352}
\textsc{Barron J.~T., Malik J.}:
\newblock Intrinsic scene properties from a single {RGB-D} image.
\newblock \emph{IEEE Trans. Pattern Anal. Mach. Intell.} (2016).

\bibitem[DHT{\etalchar{*}}00]{Debevec:2000:ARF:344779.344855}
\textsc{Debevec P., Hawkins T., Tchou C., Duiker H.-P., Sarokin W., Sagar M.}:
\newblock Acquiring the reflectance field of a human face.
\newblock In \emph{Proceedings of the 27th Annual Conference on Computer
  Graphics and Interactive Techniques} (New York, NY, USA, 2000), SIGGRAPH '00,
  ACM.

\bibitem[DRC{\etalchar{*}}15]{Duchene:2015:MII:2843519.2756549}
\textsc{Duch\^{e}ne S., Riant C., Chaurasia G., Moreno J.~L., Laffont P.-Y.,
  Popov S., Bousseau A., Drettakis G.}:
\newblock Multiview intrinsic images of outdoors scenes with an application to
  relighting.
\newblock \emph{ACM Trans. Graph.} (2015).

\bibitem[DYWG17]{DBLP:journals/corr/DongYWG17}
\textsc{Dong H., Yu S., Wu C., Guo Y.}:
\newblock Semantic image synthesis via adversarial learning.
\newblock In \emph{Proceedings of International Conference on Computer Vision
  (ICCV)} (2017).

\bibitem[GEB15]{gatys2015neural}
\textsc{Gatys L.~A., Ecker A.~S., Bethge M.}:
\newblock A neural algorithm of artistic style.
\newblock \emph{arXiv preprint arXiv:1508.06576} (2015).

\bibitem[GFT{\etalchar{*}}11]{Ghosh:2011:MFC:2070781.2024163}
\textsc{Ghosh A., Fyffe G., Tunwattanapong B., Busch J., Yu X., Debevec P.}:
\newblock Multiview face capture using polarized spherical gradient
  illumination.
\newblock \emph{ACM Trans. Graph.} (2011).

\bibitem[Gob]{Gobotree}
{Gobotree}.
\newblock \url{https://www.gobotree.com/cat/3d-people}.

\bibitem[GTB{\etalchar{*}}13]{graham2013measurement}
\textsc{Graham P., Tunwattanapong B., Busch J., Yu X., Jones A., Debevec P.,
  Ghosh A.}:
\newblock Measurement-based synthesis of facial microgeometry.
\newblock In \emph{Computer Graphics Forum} (2013).

\bibitem[GZC{\etalchar{*}}17]{DBLP:journals/corr/abs-1708-00980}
\textsc{Guo Y., Zhang J., Cai J., Jiang B., Zheng J.}:
\newblock Photo-realistic face images synthesis for learning-based fine-scale
  {3D} face reconstruction.
\newblock \emph{IEEE Trans. Pattern Anal. Mach. Intell.} (08 2017).

\bibitem[HDR]{HDRIHaven}
{HDRI Haven}.
\newblock \url{https://hdrihaven.com}.

\bibitem[HLBK18]{Huang2018}
\textsc{Huang X., Liu M.-Y., Belongie S., Kautz J.}:
\newblock Multimodal unsupervised image-to-image translation.
\newblock In \emph{Proceedings of the European Conference on Computer Vision
  (ECCV)} (2018), pp.~172--189.

\bibitem[HZRS16]{he2016deep}
\textsc{He K., Zhang X., Ren S., Sun J.}:
\newblock Deep residual learning for image recognition.
\newblock In \emph{Proceedings of the IEEE conference on computer vision and
  pattern recognition} (2016), pp.~770--778.

\bibitem[IRWM17]{10.1111/cgf.13220}
\textsc{Innamorati C., Ritschel T., Weyrich T., Mitra N.~J.}:
\newblock Decomposing single images for layered photo retouching.
\newblock \emph{Comput. Graph. Forum 36}, 4 (July 2017), 15–25.
\newblock URL: \url{https://doi.org/10.1111/cgf.13220}, \href
  {https://doi.org/10.1111/cgf.13220} {\path{doi:10.1111/cgf.13220}}.

\bibitem[IZZE17]{isola2017image}
\textsc{Isola P., Zhu J.-Y., Zhou T., Efros A.~A.}:
\newblock Image-to-image translation with conditional adversarial networks.
\newblock In \emph{Computer Vision and Pattern Recognition (CVPR), 2017 IEEE
  Conference on} (2017).

\bibitem[KW14]{kingma2013autoencoding}
\textsc{Kingma D.~P., Welling M.}:
\newblock Auto-encoding variational {Bayes}.
\newblock In \emph{Proceedings of the International Conference on Learning
  Representations (ICLR} (2014), editor, (Ed.).

\bibitem[LCY{\etalchar{*}}17]{8237510}
\textsc{{Liu} G., {Ceylan} D., {Yumer} E., {Yang} J., {Lien} J.}:
\newblock Material editing using a physically based rendering network.
\newblock In \emph{2017 IEEE International Conference on Computer Vision
  (ICCV)} (2017), pp.~2280--2288.
\newblock \href {https://doi.org/10.1109/ICCV.2017.248}
  {\path{doi:10.1109/ICCV.2017.248}}.

\bibitem[LLWT15]{liu2015faceattributes}
\textsc{Liu Z., Luo P., Wang X., Tang X.}:
\newblock Deep learning face attributes in the wild.
\newblock In \emph{Proceedings of International Conference on Computer Vision
  (ICCV)} (December 2015).

\bibitem[LMF{\etalchar{*}}19]{LeGendre2019}
\textsc{LeGendre C., Ma W.-C., Fyffe G., Flynn J., Charbonnel L., Busch J.,
  Debevec P.}:
\newblock {DeepLight}: Learning illumination for unconstrained mobile mixed
  reality.
\newblock In \emph{Proceedings of the IEEE Conference on Computer Vision and
  Pattern Recognition} (2019), pp.~5918--5928.

\bibitem[LTH{\etalchar{*}}17]{ledig2017photo}
\textsc{Ledig C., Theis L., Husz{\'a}r F., Caballero J., Cunningham A., Acosta
  A., Aitken A., Tejani A., Totz J., Wang Z., Shi W.}:
\newblock Photo-realistic single image super-resolution using a generative
  adversarial network.
\newblock In \emph{Proceedings of the IEEE Conference on Computer Vision and
  Pattern Recognition} (2017), pp.~4681--4690.

\bibitem[MGK{\etalchar{*}}19]{Meshry}
\textsc{Meshry M., Goldman D.~B., Khamis S., Hoppe H., Pandey R., Snavely N.,
  Martin-Brualla R.}:
\newblock Neural rerendering in the wild.
\newblock In \emph{Proceedings of the IEEE Conference on Computer Vision and
  Pattern Recognition} (2019), pp.~6878--6887.

\bibitem[PGC{\etalchar{*}}17]{paszke2017automatic}
\textsc{Paszke A., Gross S., Chintala S., Chanan G., Yang E., DeVito Z., Lin
  Z., Desmaison A., Antiga L., Lerer A.}:
\newblock Automatic differentiation in {PyTorch}.
\newblock In \emph{NIPS Autodiff Workshop} (2017).

\bibitem[PKA{\etalchar{*}}09]{5279762}
\textsc{{Paysan} P., {Knothe} R., {Amberg} B., {Romdhani} S., {Vetter} T.}:
\newblock A {3D} face model for pose and illumination invariant face
  recognition.
\newblock In \emph{IEEE International Conference on Advanced Video and Signal
  Based Surveillance} (2009).

\bibitem[PKD{\etalchar{*}}16]{pathak2016context}
\textsc{Pathak D., Krahenbuhl P., Donahue J., Darrell T., Efros A.~A.}:
\newblock Context encoders: Feature learning by inpainting.
\newblock In \emph{Computer Vision and Pattern Recognition (CVPR)} (2016).

\bibitem[rem]{remove.bg}
{remove.bg}.
\newblock \url{https://www.remove.bg/upload}.

\bibitem[SBT{\etalchar{*}}19]{Sun2019}
\textsc{Sun T., Barron J.~T., Tsai Y.-T., Xu Z., Yu X., Fyffe G., Rhemann C.,
  Busch J., Debevec P., Ramamoorthi R.}:
\newblock Single image portrait relighting.
\newblock \emph{ACM Transactions on Graphics (TOG) 38}, 4 (2019), 79.

\bibitem[SGK{\etalchar{*}}19]{Sengupta}
\textsc{Sengupta S., Gu J., Kim K., Liu G., Jacobs D.~W., Kautz J.}:
\newblock Neural inverse rendering of an indoor scene from a single image.
\newblock In \emph{Proceedings of International Conference on Computer Vision
  (ICCV)} (2019).

\bibitem[SHS{\etalchar{*}}17]{Shu2017}
\textsc{Shu Z., Hadap S., Shechtman E., Sunkavalli K., Paris S., Samaras D.}:
\newblock Portrait lighting transfer using a mass transport approach.
\newblock \emph{ACM Transactions on Graphics 37} (2017).

\bibitem[SKCJ18]{Sengupta2018}
\textsc{Sengupta S., Kanazawa A., Castillo C.~D., Jacobs D.~W.}:
\newblock {SfSNet}: Learning shape, reflectance and illuminance of faces in the
  wild.
\newblock In \emph{Proceedings of the IEEE Conference on Computer Vision and
  Pattern Recognition} (2018), pp.~6296--6305.

\bibitem[SPB{\etalchar{*}}14]{Shih2014}
\textsc{Shih Y.~C., Paris S., Barnes C., Freeman W.~T., Durand F.}:
\newblock Style transfer for headshot portraits.
\newblock \emph{ACM Transactions on Graphics} (2014).

\bibitem[Tel04]{Telea04}
\textsc{Telea A.}:
\newblock An image inpainting technique based on the fast marching method.
\newblock \emph{journal of graphics tools} (2004).

\bibitem[TZK{\etalchar{*}}17]{DBLP:journals/corr/TewariZK0BPT17}
\textsc{Tewari A., Zollh{\"{o}}fer M., Kim H., Garrido P., Bernard F.,
  P{\'{e}}rez P., Theobalt C.}:
\newblock {MoFA}: Model-based deep convolutional face autoencoder for
  unsupervised monocular reconstruction.
\newblock In \emph{Proceedings of the IEEE International Conference on Computer
  Vision} (2017).

\bibitem[TZN19]{thies2019deferred}
\textsc{Thies J., Zollhöfer M., Nießner M.}:
\newblock Deferred neural rendering: Image synthesis using neural textures.
\newblock \emph{ACM Transactions on Graphics} (2019).

\bibitem[XM06]{10.1145/1128923.1128974}
\textsc{Xiao X., Ma L.}:
\newblock Color transfer in correlated color space.
\newblock In \emph{Proceedings of the 2006 ACM International Conference on
  Virtual Reality Continuum and Its Applications} (2006).

\bibitem[YS19]{yu2019inverserendernet}
\textsc{Yu Y., Smith W.~A.}:
\newblock {InverseRenderNet}: Learning single image inverse rendering.
\newblock In \emph{Proceedings of the IEEE Conference on Computer Vision and
  Pattern Recognition} (2019), pp.~3155--3164.

\bibitem[ZCC18]{zheng2018t2net}
\textsc{Zheng C., Cham T.-J., Cai J.}:
\newblock {T2Net}: Synthetic-to-realistic translation for solving single-image
  depth estimation tasks.
\newblock In \emph{Proceedings of the European Conference on Computer Vision
  (ECCV)} (2018), pp.~767--783.

\bibitem[ZJ19]{Zhou}
\textsc{Zhou H., Jacobs D.~W.}:
\newblock Deep single-image portrait relighting.
\newblock In \emph{Proceedings of International Conference on Computer Vision
  (ICCV)} (2019).

\bibitem[ZPIE17]{CycleGAN2017}
\textsc{Zhu J.-Y., Park T., Isola P., Efros A.~A.}:
\newblock Unpaired image-to-image translation using cycle-consistent
  adversarial networkss.
\newblock In \emph{Computer Vision (ICCV), 2017 IEEE International Conference
  on} (2017).

\bibitem[ZZP{\etalchar{*}}17]{Zhu2017}
\textsc{Zhu J.-Y., Zhang R., Pathak D., Darrell T., Efros A.~A., Wang O.,
  Shechtman E.}:
\newblock Toward multimodal image-to-image translation.
\newblock In \emph{Advances in Neural Information Processing Systems 30}
  (2017), Curran Associates, Inc.

\end{thebibliography}
	
	
	
\end{document}